# МОДЕЛИРОВАНИЕ ПЕРЕРАСПРЕДЕЛЕНИЯ ИОННО-ИМПЛАНТИРОВАННОГО БОРА В РАЗЛИЧНЫХ УСЛОВИЯХ ПОДАВЛЕНИЯ СКОРОТЕЧНОЙ ДИФФУЗИИ

О.И. Величко, А.П. Ковалева

*Кафедра физики, Белорусский государственный университет информатики и радиоэлектроники, Беларусь, 220013 г.Минск, ул. П.Бровки, 6, тел. (+37529) 6998078, e-mail: velichkomail@gmail.com*

Результаты проведенного моделирования показывают, что при температурах отжига 800 ºC и ниже перераспределение ионно-имплантированного бора осуществляется посредством длиннопробежной миграции неравновесных межузельных атомов примеси независимо от используемого способа подавления скоротечной диффузии. Определены относительные количества атомов примеси, которые находились в межузельном положении, и значения средних длин пробега неравновесных межузельных атомов бора.

## Введение

Для создания сверхмалых областей *p*–типа проводимости при изготовлении элементов современных кремниевых интегральных микросхем обычно применяется низкоэнергетическая высокодозная имплантация ионов бора. С целью подавления скоротечной диффузии ионно-имплантированной примеси широко используется метод введения бора в слой кремния, предварительно аморфизованный внедрением более тяжелых ионов германия, который, как и кремний, является элементом IV группы (смотри, например [1,2]). Иногда, для создания аморфного слоя на поверхности полупроводника используется имплантация ионов кремния или ионов, содержащих фтор. При последующем отжиге в результате твердофазной рекристаллизации созданного таким образом аморфного слоя образуется малодефектная кристаллическая структура, а атомы бора занимают положение замещения в узлах кристаллической решетки, причем их концентрация существенно превышает предел растворимости бора в кремнии для данной температуры обработки. К сожалению, и в этом случае, при продолжении термообработки после завершения твердофазной рекристаллизации и при последующих отжигах наблюдается скоротечная диффузия бора, хотя она носит совершенно иной характер и характеризуется меньшей интенсивностью. В работах [3,4] было показано, что при температурах обработки 800 ºC и ниже скоротечная диффузия ионно-имплантированного бора в слоях, которые были аморфизованы, осуществляется по механизму длиннопробежной миграции неравновесных межузельных атомов примеси. Согласно [4] генерация межузельных атомов бора может иметь место как в результате формирования или перестройки кластеров атомов примеси, так и в результате возникновения упругих напряжений в формируемой структуре в силу отличия атомных радиусов B и Ge от атомного радиуса Si.

С целью дальнейшего подавления скоротечной диффузии может применяться еще одна имплантация ионов азота [5] или углерода [6, 7]. Было установлено, что присутствие углерода в кремнии подавляет образование остаточных нарушений [6]. Необходимым условием такого положительного эффекта является расположение атомов углерода в положении замещения [7, 8]. Если имплантированный углерод располагается в пределах аморфизованного слоя, то он оказывается в положении замещения в результате твердофазной рекристаллизации. Имплантацию углерода также можно проводить при больших значениях токов. Согласно [8] это имеет следствием рост мгновенной концентрации вакансий, что способствует переходу углерода в положение замещения. Обычно предполагается, что подавление скоротечной диффузии связано с поглощением межузельных атомов кремния в результате вытеснения атомов углерода из узлов кристаллической решетки [9,10], причем вытеснение углерода из положения замещения подтверждается рентгенодифракционным анализом [10]. Тем не менее, детальный микроскопический механизм взаимодействия атомов углерода с межузельными атомами кремния и бора до сих пор неясен. Отметим, что согласно [11] межузельные атомы кремния вытесняют из узлов кристаллической решетки любые примесные атомы. Это означает, что механизм подавления скоротечной диффузии на начальной стадии отжига после имплантации ионов азота может быть аналогичен механизму подавления при использовании атомов углерода и заключаться в уменьшении концентрации неравновесных межузельных атомов кремния и возрастании концентрации вакансий.

## Цель работы

Проведенный анализ позволяет сформулировать цель данной работы, которая заключается в моделировании перераспределения ионно-имплантированного бора в различных условиях подавления скоротечной диффузии, когда помимо создания аморфизованного слоя в приповерхностную область кремниевых подложек вводятся ионы дополнительных примесей.

## Используемая модель

В представленном исследовании будем рассматривать случай низкотемпературных обработок, когда согласно [3,4] перенос примеси должен осуществляться посредством длиннопробежной миграции неравновесных межузельных атомов бора. Как было показано в [12,13], характерным признаком реализации данного механизма переноса является протяженный "хвост" на профиле распределения концентрации примеси, который имеет вид прямой линии в случае логарифмического масштаба по оси концентраций.

## Результаты моделирования

Рассмотрим результаты моделирования экспериментальных данных [5], поскольку в этой работе использовались различные способы подавления скоротечной диффузии. Так, в работе [5] имплантация ионов бора с энергией 1 кэВ и дозой 1.5×10$^{15}$ см$^{-2}$ осуществлялась в кремниевые подложки *n*-типа проводимости, ориентации (100). Для подавления скоротечной диффузии приповерхностный слой кремния предварительно аморфизовывался имплантацией ионов Ge с энергией 15 кэВ. После имплантации бора осуществлялся термический отжиг в атмосфере азота длительностью 60 секунд при температуре 800 °C. Расчет процесса перераспределения бора, проведенный для этих условий в работе [14], показал хорошее согласие с экспериментальными данными при условии, что средняя длина пробега межузельных атомов бора $l_{AI}$ = 11 нм. Также предполагалось, что примерно 8,6 % имплантированных атомов бора временно находились в межузельном положении и затем снова стали неподвижными, заняв положение замещения. Благодаря миграции этих неравновесных межузельных атомов, произошло формирование протяженного "хвоста" на профиле распределения концентрации бора, который занимает область, расположенную в интервале от 0.02 мкм до примерно 0.1 мкм от поверхности полупроводника. Для сравнения на Рис.1 представлены результаты моделирования такого же процесса перераспределения ионно-имплантированного бора в случае, когда после имплантации ионов германия дополнительно имплантировались ионы азота с дозой 1.5×10$^{15}$ см$^{-2}$ и энергией 6 кэВ [5].

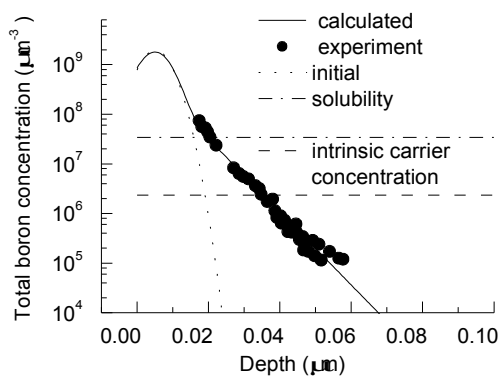

Рис.1. Рассчитанное распределение атомов ионно-имплантированного бора в кремнии, предварительно аморфизованном имплантацией ионов германия, после отжига длительностью 60 с при температуре 800 °C в условиях, когда была проведена дополнительная имплантация ионов азота с энергией 6 кэВ для более сильного подавления скоротечной диффузии. Пунктирная кривая — рассчитанное распределение бора после имплантации с энергией 1 кэВ; сплошная кривая — рассчитанное распределение бора после отжига; • — экспериментальные данные, взятые из работы Yeong et al. [5]

Как видно из рисунка, имеет место хорошее согласие результатов расчетов с экспериментальными данными. При расчете были использованы следующие значения параметров модели перераспределения имплантированного бора: **Параметры, характеризующие процесс диффузии межузельных атомов**: максимальное значение скорости генерации неравновесных межузельных бора $G_m^{AI}$ = 6.7×10$^6$ мкм$^{-3}$с$^{-1}$; средняя длина пробега межузельных атомов бора $l_{AI}$ = 5.9 нм. Концентрационный профиль распределения бора, представленный на Рис.1, был рассчитан из предположения, что примерно 24 % имплантированных атомов бора находились в межузельном положении и затем снова стали неподвижными, заняв положения замещения.

Как видно из представленных расчетов, дополнительная имплантация ионов азота приводит к существенному (почти в 2 раза) уменьшению средней длины пробега межузельных атомов бора, и уменьшению протяженности "хвоста" примерно на 0.03 мкм, что с технологической точки зрения влечет уменьшение глубины залегания *p-n* перехода. При этом значительно (в 2.8 раза) возрастает скорость генерации межузельных атомов бора. Однако, ввиду существенного уменьшения длины пробега этих атомов, большая их часть поглощается в пределах имплантированного слоя, что не приводит к увеличению глубины залегания *p-n* перехода. Мы предполагаем, что это поглощение осуществляется в результате взаимодействия с атомами азота, либо с комплексами атомов азота и германия.

Аналогичный расчет, выполненный для случая подавления скоротечной диффузии путем использования более низкой температуры обработки 750 °C, представлен на Рис.2.

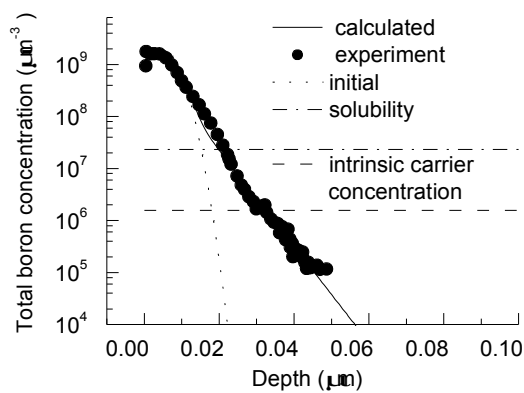

Рис.2. Рассчитанное распределение атомов ионно-имплантированного бора в кремнии, предварительно аморфизованном имплантацией ионов германия, после отжига длительностью 60 с при температуре 750 °C в условиях, когда была проведена дополнительная имплантация ионов азота с энергией 6 кэВ для более сильного подавления скоротечной диффузии

При расчете были использованы следующие значения **параметров, характеризующих процесс миграции межузельных атомов примеси**: максимальное значение скорости генерации межузельных бора $G_m^{AI}$ = 7.4×10$^6$ мкм$^{-3}$с$^{-1}$; средняя длина пробега межузельных атомов бора $l_{AI}$ = 4.7 нм. Как видно из полученных значений, снижение температуры обработки имеет следствием уменьшение средней длины пробега межузельных атомов бора, что приводит к большей крутизне профиля распределения примеси.

Наконец, на Рис.3 представлены результаты моделирования перераспределения ионно-имплантированного бора, когда для подавления скоротечной диффузии использовалось внедрение ионов BF$_2$ [15].

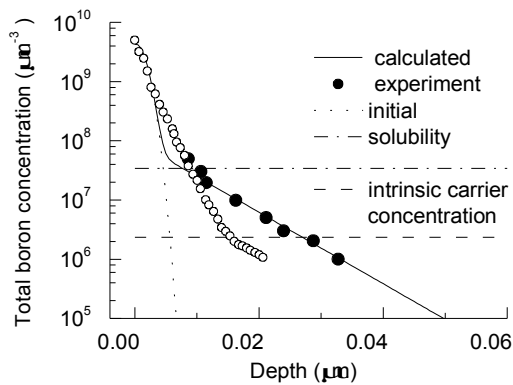

Рис.3. Рассчитанное распределение атомов бора после отжига длительностью 30 минут при температуре 800 °С кремниевых подложек, имплантированных ионами BF$_2$. Пунктирная кривая — рассчитанное распределение бора после имплантации; o — измеренный профиль распределения бора после имплантации; сплошная кривая — рассчитанное распределение бора после отжига; ● — экспериментальные данные, взятые из [15]

В работе [15] имплантация BF$_2$ осуществлялась с энергией 2.2 кэВ и дозой 1.0×10$^{15}$ см$^{-2}$. Отжиг проводился при температуре 800 °С в течение 30 минут. При расчете были использованы следующие значения **параметров, характеризующих процесс миграции межузельных атомов примеси**: усредненное по времени обработки значение скорости генерации межузельных бора в максимуме пространственного распределения $G_m^{AI}$ = 2.0×10$^5$ мкм$^{-3}$с$^{-1}$; средняя длина пробега межузельных атомов бора $l_{AI}$ = 7.2 нм. Концентрационный профиль распределения бора, представленный на Рис.3, был рассчитан из предположения, что примерно 4.6 % атомов бора находились в межузельном положении. Как видно из Рис.3, результаты расчетов хорошо согласуются с экспериментальными данными в области "хвоста" на профиле распределения бора, причем как рассчитанный, так и экспериментальный профили распределения имеют вид прямой линии. Это означает, что данная область формируется в результате длиннопробежной миграции неравновесных межузельных атомов бора.

**Заключение**

Проведено моделирование перераспределения ионно-имплантированного бора для случая подавления скоротечной диффузии посредством создания предварительно аморфизованного слоя внедрением ионов Ge и дополнительной имплантации ионов азота. Расчеты были проведены для температур обработки 750 и 800 °С при длительности отжига 60 секунд. Также исследовался случай имплантации ионов BF$_2$ и 30 минутного отжига при 800 °С. Показано, что при температурах отжига 800 °С и ниже перераспределение имплантированного бора осуществляется посредством длиннопробежной миграции неравновесных межузельных атомов примеси независимо от используемого способа подавления скоротечной диффузии. Определены значения долей атомов примеси, которые находились в межузельном положении и затем перешли в положение замещения, а также средние длины пробега неравновесных межузельных атомов бора. Наименьшее значение длины пробега межузельных атомов бора равное 4.7 нм имеет место при дополнительной имплантации ионов азота и отжиге при 750 °С.

**SIMULATION OF ION-IMPLANTED BORON REDISTRIBUTION UNDER DIFFERENT CONDITIONS OF THE TRANSIENT ENHANCED DIFFUSION SUPPRESSION**

Oleg Velichko, Alina Kavaliova

*Department of Physics, Belarusian State University of Informatics and Radioelectronics,*
*6, P.Brovki str., Minsk, 220013 Belarus*
*Tel. +375296998078, E-mail: velichkomail@gmail.com*

It has been shown by means of impurity diffusion simulation that ion-implanted boron redistribution at the annealing temperatures 800 °C and lower is governed by the long-range migration of nonequilibrium impurity interstitials regardless of the methods used for the transient enhanced diffusion suppression. The relative amounts of impurity atoms, which are being transferred to the transient interstitial position, have been determined and time-average migration lengths of nonequilibrium boron interstitials have been obtained.